\documentstyle[12pt]{article}
         \makeatletter
         \def\thefigure{\@arabic\c@figure}\def\fps@figure{tbp}
         \def\ftype@figure{1}\def\ext@figure{lof}
         \def\fnum@figure{\protect\footnotesize Fig.\ \thefigure}
         \def\thetable{\@arabic\c@table}
         \def\fps@table{tbp}\def\ftype@table{2}\def\ext@table{lot}
         \def\fnum@table{\protect\footnotesize Table \thetable}
         \def\@listI{\leftmargin\leftmargini\parsep=0pt\itemsep=0pt}
         \def\thebibliography#1{\section{References}\vspace*{-10pt}\list
          {[\arabic{enumi}]}{\settowidth\labelwidth{[#1]}\leftmargin\labelwidth
          \advance\leftmargin\labelsep
          \usecounter{enumi}}
          \def\newblock{\hskip .11em plus .33em minus .07em}
          \sloppy\clubpenalty4000\widowpenalty4000
          \sfcode`\.=1000\relax}
         
         \def\@nomath#1{\ifmmode \fi}
         \def\mmycite{\@ifnextchar [{\@tempswatrue\@mmycitex}
             {\@tempswafalse\@mmycitex[]}}
         \def\@mmycitex[#1]#2{\if@filesw\immediate%
         \write\@auxout{\string\citation{#2}}\fi
           \def\@citea{}\@mmycite{\@for\@citeb:=#2\do
             {\@citea\def\@citea{,}\@ifundefined
                {b@\@citeb}{{\bf ?}\@warning
                {Citation `\@citeb' on page \thepage \space undefined}}%
         \hbox{\csname b@\@citeb\endcsname}}}{#1}}
         \def\@mmycite#1#2{{{\scriptsize#1}\if@tempswa , #2\fi}}
         \def\mycite#1{$^{\protect\mmycite{#1}}$}
         \def\@cite#1#2{{#1\if@tempswa , #2\fi}}
         \def\thesection {\arabic{section}}
         \def\section#1{\addtocounter{section}{1}\setcounter{subsection}{0}
              \bigskip\medskip{\noindent\bf\thesection.\ #1}
              \medskip}
         \def\thesubsection {\arabic{section}.\arabic{subsection}}
         \def\subsection#1{\addtocounter{subsection}{1}
              \medskip{\noindent\thesubsection.\ #1}
              \medskip}
         \makeatother
\begin{document}
\vspace*{0.3in}
\begin{center}
  {\bf NEUTRINOS FROM PROTONEUTRON STARS: \\
A PROBE OF HOT AND DENSE MATTER } \\
  \bigskip
  \bigskip
  SANJAY REDDY AND MADAPPA PRAKASH \\
  {\em Physics Department \\
       State University of New York at Stony Brook \\
       Stony Brook, NY 11794-3800, USA }
  \bigskip
\end{center}
\smallskip
{\footnotesize
\centerline{ABSTRACT}

\begin{quotation}
\vspace{-0.10in}
Neutrino processes in dense matter play a key role in the dynamics,
deleptonization and the early cooling of hot protoneutron stars formed in the
gravitational collapse of massive stars.   Here we calculate neutrino mean free
paths from neutrino-hyperon interactions in dense matter containing hyperons.
Significant contributions to the neutrino opacity arise from scattering
involving the  $\Sigma^-$ hyperon, and absorption processes involving the
neutral and $\Sigma^-$ hyperons.   The estimates given here emphasize
the need for (a) opacities which incorporate many-body
effects in a multi-component mixture, and (b) new calculations of thermal and
leptonic evolution of protoneutron stars with neutrino transport and equations
of state with strangeness-rich matter.
\end{quotation}}

\vskip 5pt
\section{Introduction}

The general nature of the neutrino signature expected from a newly formed
neutron star (hereafter referred to as a protoneutron star) has been
theoretically predicted\mycite{BL} and confirmed by the
observations\mycite{SN87A} from supernova SN1987A.  Although neutrinos
interact weakly with matter, the high baryon densities and neutrino energies
achieved after the gravitational collapse of a massive star ($\geq 8$ solar
masses) cause the neutrinos to become trapped on the dynamical timescales of
collapse\mycite{sato,maz}.  Trapped neutrinos at the star's core have
Fermi energies $E_\nu \sim 200-300$ MeV and are primarily of the $\nu_e$
type. They escape after diffusing through the star exchanging energy with the
ambient matter, which has an entropy per baryon of order unity in units of
Boltzmann's constant.  Eventually they emerge from the star with an average
energy $\sim 10-20$ MeV and in nearly equal abundances of all three flavors,
both particle and anti-particle.

Although the composition and the equation of state of the hot protoneutron star
matter are not yet known with certainty, QCD based effective Lagrangians have
opened up intriguing possibilities. Among these is the possible existence of
matter with strangeness to baryon ratio of order unity.   Strangeness may be
precipitated either in the form of fermions, notably the $\Lambda$ and
$\Sigma^-$ hyperons, or, in the form of a Bose condensate, such as a
$K^-$-meson condensate (see Ref.~[\cite{prak}] for detailed discussion and
extensive references).  In the absence of trapped neutrinos,  strange
particles are expected to appear around $2-4$ times the nuclear matter density
of $n_0=0.16~{\rm fm}^{-3}$.  Neutrino-trapping  causes the strange particles
to appear at somewhat higher densities than in neutrino-free matter.   The
compositions shown in Figs. 1 and 2 highlight the influence of hyperons in the
neutrino trapped regime.  The results shown in these figures were
calculated\mycite{prak} using a field-theoretical model in which baryons
interact via the exchange of $\sigma , ~\omega$ and $\rho$ mesons.   With the
appearance of hyperons in matter, the electron-neutrino fraction increases with
density in contrast to the case in which matter contains nucleons and leptons
only.  A similar behavior is observed in kaon condensed matter\mycite{tpl} and
also in matter where a phase transition to quark matter occurs\mycite{pcl}.
This behavior is associated with the presence of non-leptonic negatively
charged particles in matter\mycite{prak}, such as the $\Sigma^-$ hyperon, or
$K^-$meson, or $d$ and $s$ quarks.

Keil and Janka\mycite{KJ} have recently investigated the influence of the
equation of state on the cooling and evolution of the protoneutron star.   They
find that the neutrino luminosity  depends sensitively on the composition and
on the stiffness of the  equation of state at high density.  In particular, the
influence of hyperons, which introduces a softening of the high density
equation of state, was examined.  In many cases, the protoneutron star
collapsed
to a black hole causing an abrupt cessation of  the neutrino signal.  Although
a clear distinction between the different equations of state could be achieved
on the basis of the calculated neutrino signals,  Keil and Janka conclude that
``none of the models could be considered as a good  fit of the neutron star
formed in SN 1987A''.

We note, however, that in these studies, the sensitivity of
the neutrino signals due only to the structural changes caused by the
equation of state was assessed.  In the presence of hyperons, the transport of
neutrinos is also affected due to the changes in the composition,
and additionally, to the interactions of neutrinos with strange particles.
These effects were ignored in Ref.~[\cite{KJ}].  Previous
work\mycite{max,pplp} involving neutrino interactions with hyperons  was
concerned with charged-current reactions only.  Studies of the opacities and
the transport processes of neutrinos out of the core  containing nucleons,
leptons, and in some cases,  pion condensates may be found in
Refs.~[\cite{Saw1}] through [\cite{Bur}].
%Refs.~[cite{Saw1,Tub1,LP,Lamb,Tub2,BVR,SS,Iwa2,IP,GP,BM,Good,Bru,VC,Coop,Bur}].

It is our purpose here to  study neutrino mean free paths in matter  containing
hyperons.  Specifically, we will estimate scattering and absorption mean free
paths of neutrinos when matter is under degenerate  conditions.  We find that
significant contributions to the neutrino opacity arise from  scattering
involving the  $\Sigma^-$ hyperon, and absorption processes involving the
neutral and $\Sigma^-$ hyperons.
Compared with ordinary matter, the presence of strangeness in
matter is expected to lead to an excess of $\nu_e$ neutrinos relative to other
types. Calculations of the complete thermal and leptonic evolution of a newly
formed neutron star incorporating neutrino transport and equations of state
with strangeness-rich matter will be taken up separately.   With new generation
neutrino detectors capable of recording thousands of  neutrino events, it may
be possible to distinguish between different scenarios observationally.

\begin{figure}
\vspace*{7.5in}
%\vfill
%\special{psfile=[mader.text.kw_93]phi.pro hscale=45 vscale=45
%hoffset=400 voffset=-45 angle=90}
\caption[]{\footnotesize
The composition, electron chemical potential and entropy per particle in
nucleons only matter with a lepton fraction $Y_{Le} = Y_e + Y_{\nu_e}= 0.4$,
where $Y_i=n_i/n_b$. Results are from Ref.~[\cite{prak}]. }
\end{figure}

\begin{figure}
\vspace*{7.5in}
%\special{psfile=[mader.text.kw_93]gong3.pro hscale=50 vscale=45
%hoffset=430 voffset=-10 angle=90}
\caption[]{\footnotesize
The composition, lepton chemical potentials and entropy per particle
 in strangeness-rich matter with a lepton fraction
$Y_{Le} = Y_e + Y_{\nu_e}= 0.4$, where $Y_i=n_i/n_b$.
Results are from Ref.~[\cite{prak}]. }
\end{figure}

\vskip 5pt
\section{Neutrino interactions with strange baryons }

Neutrino interactions with matter proceed via charged and neutral
current reactions. The neutral current processes contribute to
elastic scattering, and the charged current reactions result in neutrino
absorption.  The interaction Lagrangian for these reactions
is given by the Wienberg-Salam theory:
\begin{eqnarray}
{\cal L}_{int}^{nc} &=& ({G_F}/{2\sqrt{2}}) ~~l_\mu
j_z^\mu\, \qquad {\rm ~for} \qquad \nu + B \rightarrow  \nu + B \nonumber \\
{\cal L}_{int}^{cc} &=& ({G_F}/{\sqrt{2}}) ~~l_\mu^e j_w^\mu\,
\qquad {\rm for} \qquad \nu + B_1 \rightarrow  \ell^- + B_2  \,,
\end{eqnarray}
where $G_F\simeq 1.436\times 10^{-49}~{\rm erg~cm}^{-3}$ is the weak coupling
constant, $\nu$ is a neutrino, $B_1$ and $B_2$ are baryons, and $\ell^-$ is a
lepton.   The leptonic and hadronic currents appearing above are:
\begin{eqnarray}
l_\mu &=& {\overline \psi}_\nu \gamma_\mu
\left( 1 - \gamma_5 \right) \psi_\nu \nonumber\\
l_\mu^e &=& {\overline \psi}_e \gamma_\mu
\left( 1 - \gamma_5 \right) \psi_\nu \nonumber\\
j_w^\mu &=& {\overline \psi}_i \gamma^\mu
\left( g_{Vi} - g_{Ai} \gamma_5 \right) \psi_i \nonumber\\
j_z^\mu &=& {\overline \psi}_i \gamma^\mu
\left( C_{Vi} - C_{Ai} \gamma_5 \right) \psi_i \,,
\end{eqnarray}
where $i=n,p,\Lambda,\cdots$.
The neutral current process couples neutrinos of all types ($e,\mu$ and $\tau$)
to the  weak neutral hadronic current $j_z^\mu$.  The charged current processes
of interest here are electron and muon neutrinos coupled  to the charged
hadronic current $j_w^\mu$.  The vector and axial vector coupling  constants
are listed in Table 1.    Numerical values of the  parameters that best fit the
experiments are:  D=0.756 , F=0.477, $\sin^2\theta_W$=0.23 and $\sin\theta_c =
0.231$.
\vskip 0.2in

{\centerline {TABLE 1}
\vskip 0.1in
{\centerline {NEUTRAL CURRENT VECTOR AND AXIAL COUPLINGS } }
\vskip 0.1in
\begin{center}
\begin{tabular}{c|cccc}
\hline \hline
{Reaction } & $C_V$  & $C_A $ & ${\cal R}_{nc}^{(1)}$ & ${\cal R}_{nc}^{(2)}$
\\ \hline
$\nu_i + n \rightarrow \nu_i + n$ & $-1$ & $-D-F$ & 1 & 1 \\
$\nu_i + p \rightarrow \nu_i+ p$ & $ (1-4\sin^2\theta_W)$ & $D+F$ & 0.7504 &
0.8597 \\
$\nu_i + \Lambda\rightarrow \nu_i +\Lambda$ & 0& 0 & 0  & 0 \\
$\nu_i + \Sigma^-\rightarrow \nu_i +\Sigma^-$ &$ (-2+4\sin^2\theta_W)$ & $-2F$
& 0.7392  & 0.6788 \\
$\nu_i + \Sigma^+\rightarrow \nu_i +\Sigma^+$&$ (2-4\sin^2\theta_W)$ & 2F  &
0.7392 & 0.6788 \\
$\nu_i + \Sigma^0\rightarrow \nu_i +\Sigma^0$&$ 0$ & 0 & 0 & 0  \\
$\nu_i + \Xi^-\rightarrow \nu_i +\Xi^-$ & $ (-1+4\sin^2\theta_W)$ & D
& 0.2845 & 0.3238 \\
$\nu_i + \Xi^0\rightarrow \nu_i +\Xi^0$ & 1 & $-D+F$ & 0.2861 & 0.1852
\\ \hline \hline
\end{tabular}
\end{center}
\begin{quote}
NOTE.-- The quantity ${\cal R}_{nc}^{(1)} = [C_V^2+2C_A^2]_{ \nu + B
\rightarrow
\nu + B } / [C_V^2+2C_A^2]_{ \nu + n \rightarrow \nu + n }$ and
${\cal R}_{nc}^{(2)} = [C_V^2+4C_A^2]_{ \nu + B \rightarrow
\nu + B } / [C_V^2+4C_A^2]_{ \nu + n \rightarrow \nu + n }$.
\end{quote}
\vskip 5pt

\newpage
\vskip 5pt
{\centerline {TABLE 2}
\vskip 0.1in
{\centerline {CHARGED CURRENT VECTOR AND AXIAL COUPLINGS }}
\vskip 0.1in
\begin{center}
\begin{tabular}{c|ccc}
\hline \hline
{Reaction } & $g_V$ & $g_A$ & ${\cal R}_{cc}$ \\ \hline
$\nu_i + n\rightarrow e + p$ &$ 1$ & $F+D$  & 1 \\
$\nu_i + \Lambda \rightarrow e + p$ &$-\sqrt{3/2}$ &$-\sqrt{3/2}(F+D/3)$ &
0.0394 \\
$\nu_i + \Sigma^-\rightarrow  e + n$ &$-1$ &$ -(F-D)$ & 0.0125 \\
$\nu_i + \Sigma^-\rightarrow e + \Lambda $ & $ 0$ & $\sqrt{2/3}$ & 0.2055 \\
$\nu_i + \Sigma^-\rightarrow e + \Sigma^0$ &$ \sqrt{2}$ & $\sqrt{2}F$ & 0.6052
\\
$\nu_i + \Xi^-\rightarrow e + \Lambda$ &$ \sqrt{3/2}$ & $\sqrt{3/2}(F-D/3)$ &
0.0175  \\
$\nu_i + \Xi^-\rightarrow e + \Sigma^0$ &$ \sqrt{1/2}$ & $(F+D)/\sqrt{2}$ &
0.0282 \\
$\nu_i + \Xi^-\rightarrow e + \Xi^0$ &$ 1$ & $F+D$ & 0.0564 \\
$\nu_i + \Xi^0\rightarrow e + \Sigma^+$ &$ 1$ & $F-D$ & 0.2218  \\
\hline\hline
\end{tabular}
\end{center}
\begin{quote}
NOTE.-- The quantity ${\cal R}_{cc} =
[C^2(g_V^2+3g_A^2)]_{ \nu + B_1 \rightarrow
\ell + B_2 } / [C^2(g_V^2+3g_A^2)]_{ \nu + n \rightarrow \ell + p } $.
\end{quote}
\vskip 5pt
In what follows, we consider the lowest order (tree level) processes for
both elastic and absorption reactions.
The squared matrix element for neutral current reactions is given by
\begin{eqnarray}
{\overline {|{\cal M}_{12\rightarrow 34}|^2} }
= 16G_F^2~ [(C_V+C_A)^2 ~(p_1 \cdot p_2)~(p_3\cdot p_4) \nonumber \\
\hspace*{3cm} + (C_V-C_A)^2 ~(p_1\cdot p_4)~(p_2\cdot p_3) \nonumber \\
\hspace*{3.5cm} - (C_V^2-C_A^2) ~(p_2\cdot p_4)~(p_1\cdot p_3) ] \,,
\label{mes}
\end{eqnarray}
where the overline on ${\cal M}$ denotes a sum over final spins and an average
over the initial spins, and $p_i$ denotes the four momenta of particles
$i=1,4$.   The squared matrix element for the charged current
reactions is given by a similar relation,  but with the replacement
$C_V\rightarrow g_V$,   $C_A\rightarrow g_A$, and $G_F \rightarrow  G_FC$,
where $C=\cos\theta_C$ for a change of strangeness $\Delta S=0$ and
$C=\sin\theta_C$ for  $\Delta S=1$, consistent with the Cabibbo theory.

\vskip 5pt
\section{Neutrino mean free paths in degenerate matter}

We turn now to consider the mean free path of neutrinos in stellar matter
comprised of  degenerate baryons (neutrons, protons and hyperons) and leptons
under conditions of charge neutrality and chemical equilibrium.   For the
estimates below, we employ the non-relativistic approximation
for the baryons, so that the squared matrix element takes a simple form.
We treat the neutrinos only in the degenerate and non-degenerate limits.
Results for  arbitrary neutrino degeneracy and with the full matrix element in
Eq.~(\ref{mes}) will be reported elsewhere.

For elastic collisions  $1 + 2 \rightarrow 3 + 4$, where 1(3) denotes the
initial (final) neutrino and 2(4) denotes the the initial (final) baryon $B$,
the scattering relaxation  time may be calculated by linearizing the Boltzmann
equation. For small departure from equilibrium,  the inverse relaxation times
from the various components are additive; thus
\begin{eqnarray}
\frac{1}{\tau_s} =\sum_2 g_2 \int \prod_{i=2}^4 \frac {d^3p_i}{(2\pi)^3}
  ~{\cal W}_{fi}~[n_2(1-n_3)(1-n_4)~-~(1-n_2)n_3n_4] \,.
\end{eqnarray}
Above, the sum is over all species of baryons, $g_2$ is their degeneracy,
$n_i$ are the equilibrium
Fermi-Dirac distributions, and ${\cal W}_{fi}$ is the scattering rate
\begin{eqnarray}
{\cal W}_{fi} &=&  \left({\prod_{i=1}^4~2E_i}\right)^{-1}~ (2\pi)^4
\delta^4 (p_1+p_2-p_3-p_4)~
{\overline {|{\cal M}_{12\rightarrow 34}|^2} } \,.
\end{eqnarray}
The relaxation time $\tau_s$ characterizes the rate of change of
the distribution function $n_1$ due to interactions with species 2, and may be
used to define a scattering mean free path $\lambda_s = c\tau_s$.

For elastic scattering on heavy fermions, the momentum transfer is small.
Thus,
the  scattering rate may be expressed as a function of the neutrino  energy
$E_{\nu}$  and the neutrino scattering angle $\theta$.  In degenerate matter,
where the participant particles lie on their respective Fermi surfaces,  the
phase space  integration can be separated into  angle and energy
integrals\mycite{GP,Good}.  Thus for {\em degenerate neutrinos} and
when $k_BT \ll E_\nu v_{Fi}/c$, where $v_{Fi}$ is the
velocity at the Fermi surface of species $i$, the inverse relaxation time is
given by (see Ref.~[\cite{IP,Saw1}] for the result in a single component
system)
\begin{eqnarray}
\frac {1}{\tau_s} &=& \sum_i \frac {G_F^2}{12\pi^3} (C_{Vi}^2 + 2C_{Ai}^2)
m_{Bi}^2 (k_BT)^2   E_\nu
\left[\pi^2+\frac{(E_\nu-\mu_\nu)^2}{(kT)^2}\right] \,,
\label{scat1}
\end{eqnarray}
where the sum is over the baryonic components present in the system.

For {\em non-degenerate neutrinos} and when $k_BT \ll E_\nu v_{Fi}/c$, the
result of Ref.~[\cite{IP}] may be generalized to give
\begin{eqnarray}
\frac {1}{\tau_s} &=& \sum_i \frac {G_F^2}{15\pi^3} (C_{Vi}^2 + 4C_{Ai}^2)
p_{Fi}^2 E_\nu^3 \,.
\label{scat2}
\end{eqnarray}

The relaxation time for absorption through  charged current  reactions can be
calculated in a similar fashion.    When neutrinos are degenerate, absorption
on neutrons\mycite{Iwa2} and similarly on hyperons\mycite{pplp}, is
kinematically allowed.  In this case, the inverse absorption
length is given by\mycite{pplp}
\begin{eqnarray}
\frac {1}{\tau_a} &=& \sum_j \frac {G_F^2C^2}{4\pi^3} (g_{Vj}^2 + 3g_{Aj}^2)
m_{B_1} m_{B_2} (k_BT)^2   \mu_e
\left[\pi^2+\frac{(E_\nu-\mu_\nu)^2}{(kT)^2}\right] \Xi \nonumber\\
{\rm with} ~~ \Xi &=&
\theta (p_{B_2} + p_e - p_{B_1} - p_\nu) \nonumber\\
&+& {\displaystyle\frac {p_{B_2}+p_e - p_{B_1} + p_\nu}{2E_\nu} }
{}~\theta(p_\nu - |p_{B_2}+p_e-p_{B_1}|)
\label{abs1}
\end{eqnarray}
where $\theta(x)=1$ for $x\geq 1$ and zero otherwise.

When neutrinos are {\em non-degenerate} and when absorption is
kinematically allowed,  the relaxation time is
given by
\begin{eqnarray}
\frac {1}{\tau_a} = \sum_j \frac {G_F^2C^2}{4\pi^3} (g_{Vj}^2 + 3g_{Aj}^2)
m_{B_1} m_{B_2} (kT)^2 \mu_e
\left[\pi^2+\frac{E_\nu^2}{(kT)^2}\right]
\displaystyle{\frac {1}{1+e^{-E_\nu/kT} }  } \,.
\label{abs2}
\end{eqnarray}
We note that for nucleons-only matter, neutrino
absorption on single nucleons can proceed only if the proton concentration
exceeds some critical value in the range $(11-15)\%$ (see Ref.~[\cite{lpph}]).
For matter with lower proton concentrations, neutrino absorption occurs on two
nucleons.  However, as shown in Ref.~[\cite{pplp}],
neutrinos may be absorbed on single hyperons as long as the
concentration of $\Lambda$ hyperons exceeds a critical value that is less than
$3\%$ and is typically about $1\%$.

\vskip 5pt
\section{ Results and discussions }

The calculations above give the mean free path in a mixture  in which
all baryons are under degenerate conditions.  Several effects of strong
interactions must
be included before the results  in Eq.~(\ref{scat1}) through Eq.~(\ref{abs2})
may be utilized.  The renormalization of the density of states at the Fermi
surfaces results in the baryon masses $m_B$  being replaced by $m_B^* =
p_F/v_F$.   As pointed out by Iwamoto and Pethick\mycite{IP}, baryon-baryon
interactions introduce further Fermi liquid corrections. Specifically,  the
axial vector interaction is significantly suppressed, which increases the mean
free path of neutrinos in dense matter.  So far, such an analysis has been
restricted to pure neutron matter only.  In a multicomponent system, this
formalism must be extended to include correlations between the different
species present. The effect  of density correlations in the long wavelength
limit can be related to the compressibility of the system\mycite{Saw1}.   In
addition, the medium dependence of $g_A$ must also be considered.   For
example, it has been argued\mycite{WRB} that $|g_A|$ is quenched for nucleons
in a medium.  Whether $g_A$ for hyperons is similarly affected is not known and
is worth studying.  Finally, depending on the momentum transfers involved,
final-state interactions  may modify the weak-interaction matrix element.    Of
the many corrections mentioned above, those due to the  effective mass are the
easiest to incorporate, since it would be  contained even in a mean field
description of the equation of state.  Pending a more complete analysis, we
consider below  the modifications introduced by the multicomponent nature of
the system incorporating only the effective mass corrections.

When neutrinos are  degenerate, the relative abundances of the individual
components do not play a significant  role in determining the total mean free
path. However, the extent to which the neutrino mean free path is altered by
the presence of hyperons depends on the number of hyperonic species present.
This may be illustrated by using the results for
${\cal R}^{(1)}_{nc} = [C_V^2+2C_A^2]_{ \nu + B \rightarrow
\nu + B } / [C_V^2+2C_A^2]_{ \nu + n \rightarrow \nu + n }$
and ${\cal R}_{cc} = [C^2(g_V^2+3g_A^2)]_{ \nu + B_1 \rightarrow
\ell + B_2 } / [C^2(g_V^2+3g_A^2)]_{ \nu + n \rightarrow \ell + p} $
listed in Tables 1 and 2. For example, in matter containing $\Lambda,\Sigma^-$
and $\Sigma^0$ hyperons,  the scattering mean free path is reduced by about
$30-50\%$ from its value in nucleons only matter.   This reduction is
achieved mainly by scattering on the $\Sigma^-$ hyperon, since the $\Lambda$
and $\Sigma^0$ hyperons do  not contribute to scattering in lowest order.
Similarly, up to $50\%$ reduction may be expected from absorption reactions.
Again, reactions involving the $\Sigma^-$ hyperon, particularly the one leading
to the  $\Sigma^0$ hyperon, give the largest contribution.

The relative importance of the scattering and absorption reactions by
degenerate neutrinos may be inferred by noting that
\begin{eqnarray}
\frac{\lambda_{a}}{\lambda_{s}}&=&\frac{1}{3}\displaystyle
 \frac{E_\nu}{\mu_e}~
\left(\frac{\sum_i~(C_{Vi}^2+2C_{Ai}^2)m^{*^2}_{B_i} }
{\sum_j~C^2(g_{Vj}^2+3g_{Aj}^2)m^*_{B_1}m^*_{B_2} } \right) \,.
\end{eqnarray}
For a fixed lepton fraction and for neutrinos of energy $E_\nu \sim \mu_\nu$,
the factor $\mu_\nu/\mu_e \leq 1$ (see Figs. 1 and 2).  Inasmuch as the baryon
effective masses are all similar in magnitude, the factor in the parenthesis
above may be approximated by the ratio of the coupling constants.   From the
results in Tables 1 and 2, it is easy to verify that the factor containing the
coupling constants is of order unity (but generally less than unity) and is not
very sensitive to the number of hyperonic species present. We thus arrive at
the result that the absorption reactions dominate over the scattering reactions
by a factor of about three to five, even in the presence of hyperons.  This
conclusion is not affected by the inclusion of the baryon effective masses in
the calculation of  $\lambda_a/\lambda_s$.

For non-degenerate neutrinos, and when $k_BT \ll E_\nu v_{Fi}/c$,
\begin{eqnarray}
\frac{\lambda_{a}}{\lambda_{s}} = \displaystyle
 \frac{4}{15}~
\left(\frac{\sum_i~(C_{Vi}^2+4C_{Ai}^2) p_{Fi}^2E_\nu^3  (1+e^{-E_\nu/kT}) }
{\sum_j C^2(g_{Vj}^2 + 3g_{Aj}^2)
m^*_{B_1} m^*_{B_2} (kT)^2 \mu_e
\left[\pi^2+ {\displaystyle\frac{E_\nu^2}{(kT)^2}} \right] } \right) \,.
\end{eqnarray}
In this case, the abundances of the various particles play a significant role
in determining whether or not the absorption reactions dominate over the
scattering reactions.  Some insight may be gained by examining the
concentrations in Figs. 1 and 2, and also the ratio ${\cal R}^{(2)}_{nc} =
[C_V^2+4C_A^2]_{ \nu + B \rightarrow  \nu + B } / [C_V^2+4C_A^2]_{ \nu + n
\rightarrow \nu + n }$ listed in Table 2.

\vskip 5pt
\section{ Outlook }

Neutrino signals in terrestrial detectors offer a means to determine the
composition and the equation of state of dense matter.  Many  calculations of
the composition of dense matter have indicated that  strangeness-rich matter
should be present in the core of neutron stars.   Possible candidates for
strangeness include hyperons, a $K^-$ condensate, and quark matter containing
$s$-quarks.  Neutrino opacities for strange quark matter have been calculated
previously\mycite{Iwa2}.  In this work, we have identified the relevant
neutrino-hyperon scattering and absorption reactions that are important new
sources of opacity.   Much work remains to be done, however.  Many-body
effects which may introduce additional correlations in a mixture are worth
studying.  Full simulations, of the type carried out in Refs.~[\cite{BL,KJ}]
including possible compositions and the appropriate neutrino opacities, will be
taken up separately.

\vskip 5pt
\section{Acknowledgements}

This work was supported by the U.S. Department of Energy under contract number
DOE/DE-FG02-88ER-40388.  We thank Jim Lattimer for helpful discussions.

\end{document}